\documentclass[preprint2]{aastex}
\usepackage{graphicx}
\advance \topmargin by -\headheight
\advance \topmargin by -\headsep
\evensidemargin \oddsidemargin
\newcommand{\microns}{\micron}

\shorttitle{Mid-IR Disk Around WL16}
\begin{document}

\title{Structure of the Mid-Infrared Emitting Disk Around WL16}

\author{Michael E. Ressler\altaffilmark{1}}

\affil{Jet Propulsion Laboratory, California Institute of Technology,\\
 4800 Oak Grove Drive, Pasadena, CA 91109\vskip2.5ex}

\and
\author{Mary Barsony\altaffilmark{1,2,3}}

\affil{Space Science Institute\\
 3100 Marine Street, Suite A353, Boulder, CO 80303--1058}

\altaffiltext{1}{Visiting Astronomer at the W.M. Keck Observatory, which is
operated as a scientific partnership among the California Institute of
Technology, the University of California and the National Aeronautics and
Space Administration. The Observatory was made possible by the generous
financial support of the W.M. Keck Foundation.}

\altaffiltext{2}{NSF POWRE Visiting Professor, Physics Department, Harvey
Mudd College, Claremont, CA 91711} 

\altaffiltext{3}{NSF CAREER Award Recipient}

\begin{abstract}
WL16 is a unique member of the embedded young stellar population in the
nearby $\rho$~Ophiuchi cloud core: its extended, high surface brightness
disk is visible only at mid-infrared wavelengths. We present
diffraction-limited images, from 7.9 to 24.5~\microns, of WL16 acquired at
the Keck II telescope. We take advantage of the $\sim $0\farcs3 angular
resolution of the mid-infrared images to derive physical parameters for the
central object by self-consistently combining them with available
near-infrared spectroscopy, point-spread-function fit photometry, and
pre-main-sequence evolutionary tracks. We find the central star to be a
250~L$_{\sun}$, 4~M$_{\sun}$, Herbig Ae star, seen through foreground
material of the $\rho$~Oph cloud core that provides an extinction of
$A_{V}=31\pm 1$ magnitudes. WL16's disk is detected through all nine
observed passbands, not only those four which sample PAH emission features.
We confirm, therefore, that the emitting particles are composed of both
polycyclic aromatic hydrocarbons (PAHs) and very small (5--100~\AA)
graphitic grains (VSGs). The disk size as observed through the four PAH
filters is 7\arcsec$\times $3\farcs5 , corresponding to a disk diameter of
$\sim $~900~AU. The disk's major axis is at a position angle of $60^{\circ
}\pm 2^{\circ }$ and viewed at an inclination angle of $62.2^{\circ }\pm
0.4^{\circ }$ to our line-of-sight. Our derived inclination angle is in
excellent agreement with the previously inferred inclination for the inner
disk ($R\leq 30$ R$_{\sun }$) from kinematic modeling of the near-infrared
spectral lines of CO. We can distinguish structure within the PAH disk at
unprecedented resolution. We confirm a resolved (1\farcs5 diameter) core
component at 7.9 and 8.8~\microns, due to emission from positively charged
PAHs. An enhancement in the emission at 12.5~\microns\ at the disk's edges
is found for the first time, and signals the presence of larger ($\geq
50$--80 carbon atoms) PAHs and/or more hydrogenated PAHs than those found
in the bulk of the disk. We find a disk asymmetry, observed at all nine
mid-infrared wavelengths, at projected radii 1\arcsec--2\farcs5
(corresponding to 125~AU $\leq r\leq $ 300~AU) from the central source.
\end{abstract}


\section{Introduction}

WL16 was first discovered via near-infrared photometry \citep{wil83} to be
a member of the deeply embedded cluster currently forming in the highest
extinction core of the nearby $\rho$~Ophiuchi cloud (\citealp{bar97}, and
references therein); it is undetectable at optical wavelengths. WL16 was
identified as a few $\times 10^{5}$~yr old, late-stage protostar based on
its rising infrared (2--60~\micron) spectral slope, and was assumed to
consist of a remnant infalling dust and gas envelope surrounding a central
protostar$+$disk system \citep{lad87,ada87,wil89}.

Subsequent observations conflicted with the classification of WL16 as a
late-stage protostar, however. WL16 lacks, by two orders of magnitude, the
amount of millimeter flux expected from a low-mass young stellar object
(YSO) at such an early evolutionary stage, and shows no evidence of a
molecular outflow \citep{cab91,and94}. WL16 is the only source in Ophiuchus
known to exhibit the entire suite of polycyclic aromatic hydrocarbon (PAH)
emission features \citep{tan90,han92,deu95,nat95}. It has become clear that
PAH emission accounts for the appearance of WL16's mid-infrared spectral
features rather than what had previously been interpreted as a deep
silicate absorption feature \citep{han92}, such as is generally associated
with the presence of a cool circumstellar envelope in late-stage, low-mass
($\leq $ 2 M$_{\sun }$) protostars.

WL16 is also the first YSO in which the presence of a hot (1000K $\leq $ T
$\leq $ 6000K), dense, neutral, gaseous, Keplerian disk was inferred via
detailed modeling of its 2.3 \micron\ CO overtone emission line shapes
\citep{den91,car93,cha93,cha95}. The inner disk (3--30~R$_{\sun }$) around
WL16 is modeled as having a high inclination angle ($>48^{\circ }$) with a
fairly low accretion rate from the disk to the star of $\dot{M}\sim 2\times
10^{-7}$ M$_{\sun }$ yr$^{-1}$\citep{na96a}. Velocity-resolved spectroscopy
of the Br$\gamma $ hydrogen line toward WL16 provides independent evidence
for the low accretion rate infalling gas at spatial scales smaller than
that of the inner disk radius \citep{na96b,har94}.

What makes WL16 truly unique, however, is the large extent (8\arcsec\ along
its major axis) of its elliptically-shaped mid-infrared emission, as
revealed by previous imaging studies \citep{deu95,eme96,moo98}.
Spatially-resolved mid-infrared spectroscopy of this elongated dust
structure has confirmed the presence of both very small grains (VSG's) and
polycyclic aromatic hydrocarbons (or PAH's) throughout \citep{dev98}.

The presence of mid-infrared emission out to such large extents from a
central source is not expected from ``classical'' or ``standard
equilibrium-emitting'' silicate and amorphous carbon or ``MRN'' grains
\citep{mrn77}. These ``standard'' grains range in size from 10~nm to $\geq
250$~nm, have a size distribution, $n(a)\propto a^{-3.5}$, and are
generally assumed to be large enough to come into equilibrium with the
surrounding radiation field (although see \citealt{lyn00} for a thorough
discussion of the shortcomings attendant upon this assumption).

By contrast, the so-called graphitic ``Very Small Grains'' (VSGs), which
range in size from 0.4~nm $\leq \, \, a\, \, \leq $ 10~nm with a size
distribution $n(a)\propto a^{-4}$ \citep{dra85} and the PAHs, which contain
25--40 carbon atoms, or PAH clusters, containing 50--500 carbon atoms
\citep{sie92}, are heated and emit in a non-equilibrium fashion. They are
therefore capable of emitting in the mid-infrared, at large distances from
a central illuminating source at (seemingly) ``high temperatures'' not in
equilibrium with the surrounding radiation field.

Detailed radiative transfer modeling of the spectral energy distribution of
physically realistic, spherically symmetric dust shells surrounding Herbig
Ae stars shows that whereas the mid-infrared ($\sim $ 2--20 \micron)
emission from ``classical'' grains is negligible, the continuum emission
from VSGs in this wavelength range accounts for all of the observed
mid-infrared excesses (above photospheric) for these sources \citep{nat93}.
These authors also note that VSGs do not emit in the J band
(1.25~\microns), and only very little in the H band (1.65~\microns) for the
same models in which VSGs contribute essentially all of the observed
mid-infrared continuum emission. Therefore, the slope of the spectrum in
the 1.25--1.65 \micron\ continuum is effectively determined by the
underlying stellar radiation field and of the absorption and scattering
properties of the ``standard'' grains, even in the presence of VSGs.

Whereas VSGs essentially determine the mid-infrared continuum emission in
the circumstellar environments of Herbig Ae/Be stars \citep{her94}, the
effect of adding PAH emitters results in emission into several,
well-defined solid-state features, with minimal change in the underlying
shape of the mid-infrared continuum, as compared with models computed in
the absence of PAHs. When a UV/visible photon from the central source
interacts with a PAH lattice, the photon energy is redistributed into
stretching and bending modes of C$-$H and C$=$C bonds; these modes re-emit
energy in broad, solid-state features. These well-known PAH features are
centered at 3.29~\microns\ (the $v=1\rightarrow 0$ C$-$H stretch), at 6.2
and 7.7~\microns\ (C$=$C stretch modes), 8.6~\microns\ (C$-$H in-plane
bend), 11.3~\microns\ (C$-$H out-of-plane bend), and 12.5~\microns\ (C$-$H
bond resonance) \citep{tok91,sie92,bro93,han95}. The region between 10 and
20~\microns\ is dominated by a resonance between 10.5~\micron\ $\leq
\lambda \leq 13.5$~\microns\ (encompassing the 11.3 and 12.5~\micron\
features) due to C$-$H out-of-plane bending. Although the approximate
wavelength of one resonance is about 11.3~\microns, due to a mono-C$-$H
vibration, this wavelength can actually be longer, depending on
hydrogenation (the number of H atoms attached to the edge carbon rings).
Some continuum emission is also produced by the overlapping wings of these
PAH features.

The excitation of mid-infrared emission by PAHs and VSGs out to such large
physical distances from the central object as is observed in WL16, requires
both a fairly luminous source of optical/UV photons and optically thin
paths from the central object to the mid-IR emitting regions. The
uniqueness of the mid-IR emitting structure observed in WL16 amongst all of
the known embedded sources in $\rho$~Oph, combined with the fact that many
intermediate mass (2~M$_{\sun }\leq M\leq 8$~M$_{\sun }$) YSO's show
PAH-feature emission \citep{wat98}, leads one to re-consider the central
source's properties.

In order to shed further light on both the nature of the central object and
its surrounding circumstellar material, we have obtained images of WL16 at
nine distinct wavelengths---through filters which variously either include
or exclude PAH feature emission and span the 8--25~\micron\ atmospheric
windows---all at unprecedentedly high spatial resolution ($\sim $
0\farcs3). In addition, we have made use of previously published
near-infrared photometry, recently published spectroscopy, and up-to-date
pre-main-sequence tracks for intermediate stellar masses, to re-examine the
nature of the central object.

\section{Observations}

We observed WL16 with JPL's mid-infrared camera, MIRLIN \citep{res94}, at
the visitor port of the Keck II telescope on UT 14~March 1998 and
27~January 1999. The sky was clear and dry on both nights ($\tau
_{225GHz}\sim 0.06$ and $\sim 0.04$, respectively). MIRLIN employs a Boeing
HF--16, 128$\times $128 pixel, Si:As impurity band conductor detector
array, and produces a plate scale of 0\farcs138 per pixel (17\farcs5
field-of-view) with the f/40 chopping secondary mirror at Keck II.

\begin{figure}[!tb]
\includegraphics[width=\columnwidth]{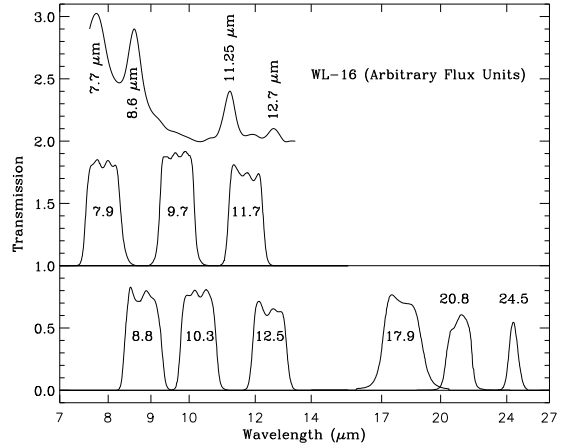}
\caption{MIRLIN filter transmission curves along with a
continuum-subtracted spectrum of WL16 \citep{dev98}. The filter curves are
reproduced from the manufacturer's data taken at 77~K. The filters at 7.9,
9.7, and 11.7~\microns\ have had their baselines shifted by one for
clarity. The 9.7 and 10.3~\micron\ filters sample the ``continuum''
reasonably cleanly, while the 7.9, 8.8, 11.7, and 12.5~\micron\ filters
each sample a PAH feature.\label{fig:filt}}
\end{figure}

Background subtraction was performed by chopping the telescope secondary
mirror 8\arcsec\ in a north-south direction, then nodding the entire
telescope 30\arcsec\ east-west, completely off the source, in order to
remove residual differences. Observations were performed through the six
10~\micron\ ``silicate'' filters (7.9, 8.8, 9.7, 10.3, 11.7, and
12.5~\microns) and the 17.9~\micron\ filter on 14~March 1998, and through
the 20.8~\micron\ and 24.5~\micron\ filters on 27~January 1999. All of
these filters have a passband of $\Delta \lambda /\lambda \approx 10$\%,
except for 24.5 \microns\ where $\Delta \lambda /\lambda \approx 3$\%.
Plots of the 9 filter passbands are shown in Figure~\ref{fig:filt}, along
with the continuum-subtracted mid-infrared spectrum of WL16 \citep{dev98},
to indicate how the PAH emission features were sampled. Note that the
well-known, broad silicate features, due to the Si--O stretching and
bending modes, at 9.7 and 18~\microns, respectively, were also sampled.

\begin{figure*}[!tb]
\includegraphics[width=\textwidth]{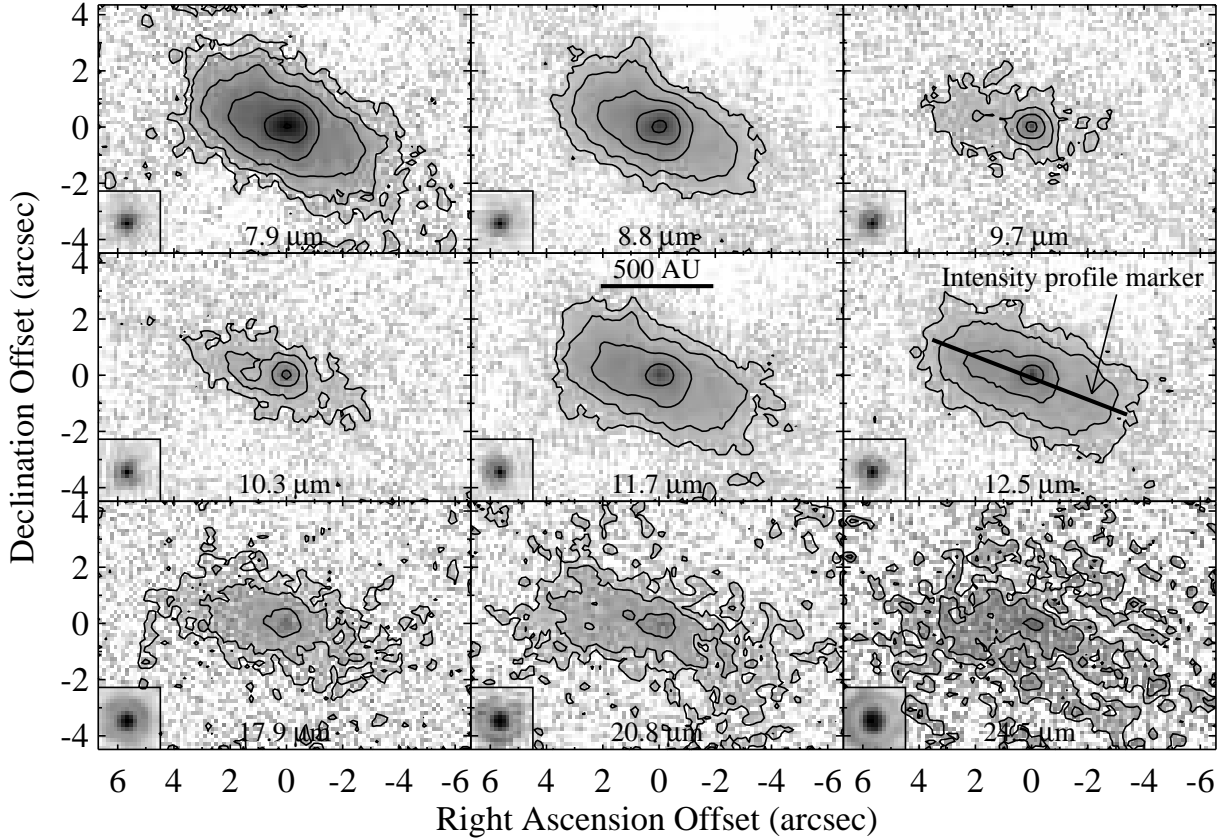}
\caption{Images of WL16 at all observed mid-infrared wavelengths. North is
up, east is left for all images. The pixel scale is 0.138 arcsec/pix,
giving the frames a roughly $13\arcsec \times 9\arcsec $ field of view. The
PSF calibrators were $\beta $ Leo at $\lambda <13$~\microns, and $\alpha $
Hya at longer wavelengths, and are shown in the insets at each wavelength.
Contours in each panel represent the flux density in Jy~arcsec$^{-2}$, and
are spaced by one magnitude. The lowest contour level in all images starts
at 50~mJy~arcsec$^{-2}$; subsequent contour levels are at 0.125, 0.315,
0.792, 1.9, and 5.0~Jy~arcsec$^{-2}$. The FWHMs of the PSFs at each
wavelength are the diffraction-limited values of 0\farcs17 at 7.9~\microns,
0\farcs19 at 8.8~\microns, 0\farcs21 at 9.7~\microns, 0\farcs22 at
10.3~\microns, 0\farcs25 at 11.7~\microns, 0\farcs27 at 12.5~\microns,
0\farcs39 at 17.9~\microns, 0\farcs45 at 20.8~\microns, and 0\farcs53 at
24.5~\microns. The ``intensity profile marker'' in the 12.5 \micron\ image
shows the axis along which the brightness profiles of Figure~\ref{fig:cuts}
are plotted.\label{fig:images}}
\end{figure*}

Total on-source integration times through each of the ``silicate'' filters
were approximately 30~seconds---200 coadded chop pairs of roughly 80~msec
duration in each of the two beams. Total on-source integration times
through the longer wavelength filters were 2.0, 2.7, and 2.4~minutes
through the 17.9, 20.8, and 24.5~\micron\ filters, respectively. The
primary photometric standard and point spread function calibrator was the
A3V star, $\beta $~Leo, which has a magnitude ranging from 1.91 to 1.84
from 7--13~\microns. The calibrator at 17.9 and 20.8~\microns\ was $\alpha
$~Hya, a K3III star with a magnitude of $-$1.49; the standard at
24.5~\microns\ was $\beta $~Lib, a 2.84 magnitude B8V star. Consistency
checks were performed with $\alpha $~CMa (mag = $-$1.39) and $\sigma $~Sco
(mag $\sim $ 2.40); the latter proved to be an easily-resolved binary with
0\farcs45 separation. Atmospheric extinction was corrected by observing the
calibrators at several different airmasses. The resulting coefficients were
found to be quite low ($<$ 0.1--0.2 magnitudes per airmass).

\begin{figure*}[!tb]
\includegraphics[width=\textwidth]{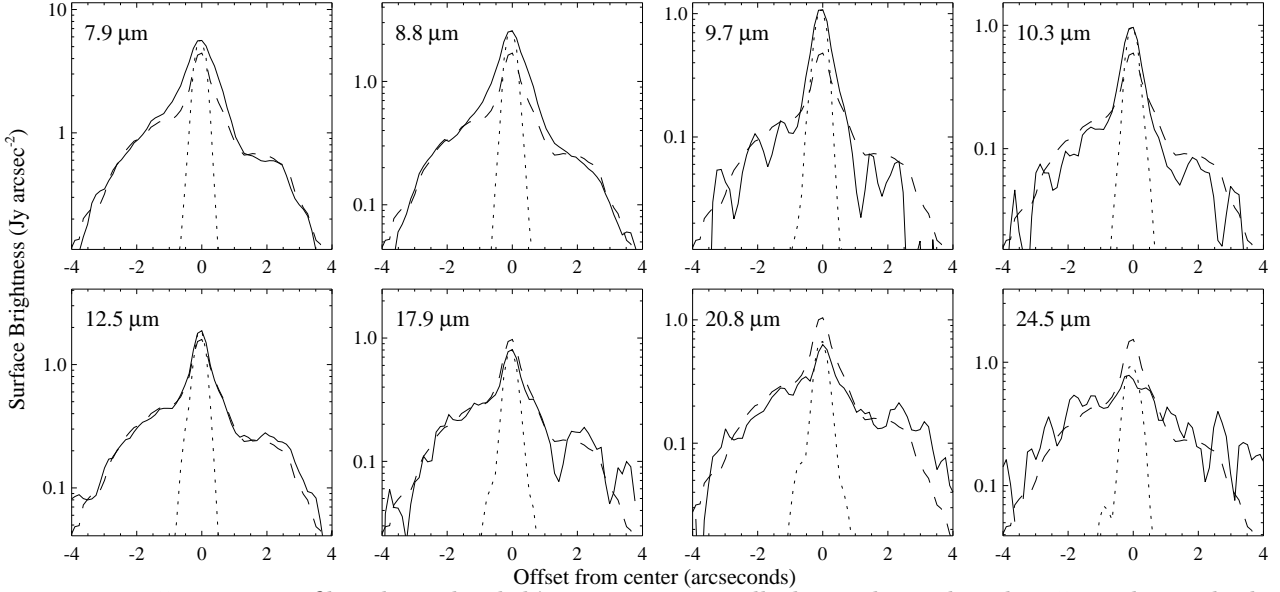}
\caption{Intensity profiles along the disk's major axis at all observed
wavelengths. In each panel, the solid curve is the intensity profile at the
indicated wavelength, the dotted curve is the PSF profile at that
wavelength, and the dashed curve is the scaled profile at 11.7~\microns
(normalized at a point in the ``shoulder'' 1.5 arcsec northeast from the
center).\label{fig:cuts}}
\end{figure*}

\section{Results}

\subsection{Mid-Infrared Images}

The mid-infrared images of WL16 are presented in Figure~\ref{fig:images}.
It is clear that the circumstellar structure surrounding WL16 appears
brighter with respect to the central source in the filters that encompass
the PAH features, relative to its appearance in the filters that exclude
PAH feature emission over the 7--13 \micron\ region. Nevertheless, the
mid-infrared emission is still well-resolved even in the filters that
exclude PAH feature emission.

The extent of the emission derived from the images that include the PAH
features is 7\arcsec$\times $3\farcs5 at 1\% of the peak level. This
projected source size corresponds to 880 $\times $ 440~AU for an assumed
distance of $d=125$~pc \citep{deg89,deg92,knu98}. The position angle of the
major axis, measured east from north, is $60^{\circ }\pm 2^{\circ }$.

Intensity profiles along the major axis of WL16 (indicated on the 12.5
\micron\ image of Figure~\ref{fig:images}) are presented for each
wavelength in Figure~\ref{fig:cuts}. The central, bright core is resolved
at the shortest (7.9 and 8.8~\micron) wavelengths, but is unresolved at
wavelengths longer than this. We also find that the contrast of the central
peak to the disk emission varies with wavelength (see Col.~5 of Table~1),
with the highest PSF/disk contrast occurring in the 9.7 and 10.3 \micron\
images and the lowest PSF/disk contrast in the 24.5~\micron\ image.

\subsection{Mid-Infrared Color Ratios}

\begin{figure*}[!p]
\includegraphics[width=\textwidth]{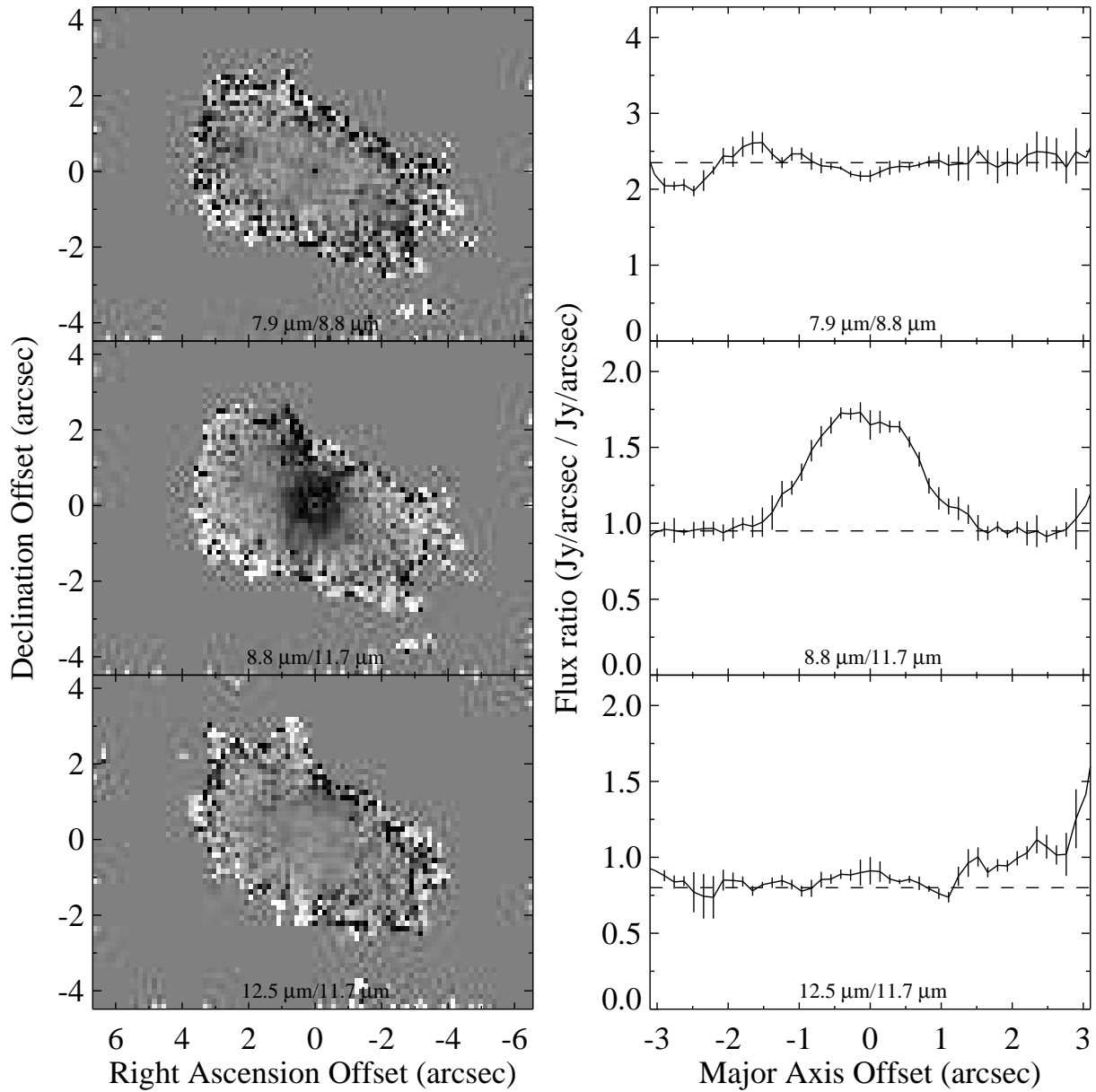}
\caption{Color ratio images of WL16 and corresponding cross-cuts taken
along the disk's major axis for selected wavelength pairs: Top panels:
7.9/8.8~\microns, Middle panels: 8.8/11.7~\microns, and Bottom panels:
12.5/11.7~\microns. North is up, east is left for each image. The pixel
scale is 0.138 arcsec/pix, giving a 13\arcsec$\times $9\arcsec\ field of
view. The images were divided on a pixel-by-pixel basis and are presented
unsmoothed. On the right hand side, we plot the corresponding flux ratios
along the major axis as a function of angular distance from the central
source. The ratio images were rotated by 30$^{\circ }$, then the median and
standard deviation were computed from a 7 pixel (1.0 arcsec) wide strip
centered on the major axis and plotted accordingly.\label{fig:gradients}}
\end{figure*}

In Figure~\ref{fig:gradients}, we present three mid-infrared color ratio
images of the WL16 disk: the 7.9/8.8~\micron\ flux ratios in the top left
panel, the 8.8/11.7~\micron\ flux ratios in the middle left panel, and the
12.5/11.7~\micron\ flux ratios in the bottom left panel. We find no
significant color gradient in the entire 7.9/8.8~\micron\ flux ratio image.
This is in contrast to the other two flux ratio images presented in
Figure~\ref{fig:gradients}. The 8.8/11.7~\micron\ image (and in the
7.9/11.7~\micron\ image not shown here) shows significantly enhanced values
within the inner $\sim$~1\farcs5 diameter ``core'' (see also the
corresponding panels of Figure~\ref{fig:cuts}). This inner core flux ratio
enhancement is more quantitatively illustrated in the corresponding
cross-cut panel of Figure~\ref{fig:gradients}, which shows the
8.8/11.7~\micron\ flux ratio at a peak value of $\sim $ 1.75 within the
inner core, declining to a relatively constant value of $\sim$~0.95
outside this core. By contrast, the 12.5/11.7~\micron\ flux ratio image
shows little variation in the inner ($\leq 2$\arcsec) regions; but shows a
small enhancement towards the disk's edges. This flux ratio rises from a
value of $\approx $ 0.85 in the inner regions to values $\geq 1.1$ at the
disk's outer edges.

\subsection{Photometry}

We present new mid-infrared and far-infrared photometry for WL16 in
Table~1. To compare our photometry with previous ground-based results
(\emph{e.g.} \citealt{lad84,moo98}), we have performed false aperture
photometry with an 8\arcsec\ diameter software aperture. In general, the
agreement between our 8-arcsec aperture photometry and that of
\citeauthor{lad84} is quite good in the mid-infrared. The random errors in
our flux measurements are about 3\% at 8--13 \microns, 7\% at 17.9 and
20.8~\microns, and 30\% at 24.5 \microns. Our fluxes agree with those
previously published to 2$\sigma $ at 8.8 \microns, 3$\sigma $ at 10.3
\microns, and $<$ 1$\sigma $ at 11.7 and 12.5 \microns. We find some
discrepancy, however, between our 9.7 \micron\ flux (2.2 Jy) and the
previously published 9.7 \micron\ flux (1.6 Jy) of \citeauthor{lad84},
though this may not be significant due to the strong, telluric ozone
absorption feature contained within this band.

We also performed point spread function fit (PSF-fit) photometry of the
central peak in each image of WL16 to better determine the flux from just
the central source, since the spatial resolution of the mid-infrared images
allows a clean separation of the central peak from the surrounding
nebulosity. To accomplish this, a PSF image derived from observed standard
stars at each wavelength was shifted, scaled, and offset to fit the central
source in WL16. This same method was applied to near-infrared images of
WL16 to derive PSF-fit near-infrared photometry, as well \citep{res92}. At
all wavelengths, the PSF-fit flux contains just photospheric light plus any
warm dust within the PSF ($<40$~AU). Values of these PSF-fit fluxes are
also tabulated in Table~1. \begin{deluxetable}{cccccc}
\tablewidth{0pt}
\tablenum{1}
\tablecaption{Photometry of WL16 \label{tbl-1}}
\tablehead{
\colhead{Wavelength} & \colhead{PSF Fitting} & \colhead{Aperture Phot.} & 
\colhead{Prev. Publ.} & \colhead{Disk/PSF Ratio} & \colhead{Refs.} \\\colhead{\(\lambda_{0}\)} & \colhead{Flux} & \colhead{Flux\(^{\mathrm {a}}\)} & 
\colhead{Flux} & 
\colhead{} \\\colhead{(\(\mu\)m)} & \colhead{(Jy)}  &\colhead{(Jy)} &
\colhead{(Jy)} & \colhead{(Aper.-PSF)/PSF} & \colhead{}  }
\startdata
 1.2 &  0.0037 &  \ldots     &0.0040     & & (1,2) \\
 1.65 & 0.059  &  \ldots     &0.071      & & (1,2) \\       
 2.2 &  0.45   &  \ldots     &0.48       & & (1,2) \\
 3.4 &  1.17   &  \ldots     &1.36       & & (1,2) \\
 3.8 &  1.22   &  \ldots     & \ldots    & & (1) \\
 4.8 &  2.18   &  \ldots     &2.02       & & (1,3) \\
 7.9 &  1.49   & 19.13       & \ldots    & 11.8&  \ldots \\
 8.8 &  0.65   &  7.89       &8.3        & 11.1&  (3) \\
 9.7 &  0.35   &  2.18       &1.6        & \ 5.2&  (3) \\
 10.3 & 0.30   &  1.97       &1.8        & \ 5.5&  (3) \\
 11.7 & 0.49   &  7.07       &7.1        & 13.4&  (3) \\
 12.5 &  0.57  &  6.30        &6.25      & 10.1&  (3) \\
 17.9 &  0.28  &  4.53        & \ldots   & 15.2&  \ldots \\
 20.8 &  0.25  &  3.19        &4.8\(^{\mathrm {b}}\)       & 11.8 & (3) \\
 24.5 &  0.19  &  5.67        & \ldots   & 28.8&  \ldots \\
 12   & \ldots & 16.0 (HIRES) &  18.8    & & (4) \\
 25   & \ldots & 14.4 (HIRES) & \(<\)40    & & (4) \\
 60   & \ldots & 129 (HIRES)  & 202      & & (4) \\
100   & \ldots & 223 (HIRES)  & \ldots   & & (4) \\
1300  & \ldots & \ldots       & \(<\)0.006 & & (5) \\\tablenotetext{a}{All quoted fluxes are for an 8\(^{\prime\prime}\) diameter
aperture, except for HIRES fluxes which were determined from a
45\(^{\prime\prime}\) diameter aperture after local background subtraction.}
\tablenotetext{b}{Broadband Q measurement---\(\Delta\lambda/\lambda \sim\) 40\%}
\tablerefs{(1)~Ressler (1992), (2)~Wilking \& Lada (1983), (3)~Lada \& Wilking
(1984), (4)~Young, Lada, \& Wilking (1986), (5)~Motte, Andr\'{e}, \& Neri (1998)}
\enddata
\end{deluxetable}

\begin{figure}[!htb]
\includegraphics[width=\columnwidth]{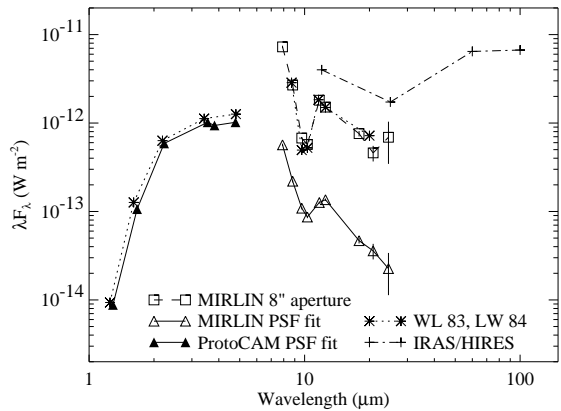}
\caption{The observed spectral energy distributions (SEDs) of WL16.
Triangles connected by a solid curve represent measurements derived from
PSF fitting (filled triangles are near-infrared data from \citealt{res92};
open triangles are mid-infrared data from this work). Fluxes derived from
our aperture photometry (using 8-arcsec software apertures) are shown by
open squares connected with a dashed curve. Data from \citet{wil83} and
\citet{lad84} are plotted with asterisks connected by a dotted curve. (The
curve is broken between 4.8 and 8.8 \microns\ since they obtained no data
at 7.9 \microns.) Crosses connected by the dot-dashed lines represent newly
determined \emph{IRAS} HIRES fluxes (this work). The dramatically
increasing flux levels with increasing aperture observed at all
mid-infrared wavelengths, including bandpasses that exclude PAH feature
emission, attest to the presence of both spatially extended PAHs and VSGs
in the circumstellar environment of WL16.\label{fig:sed}}
\end{figure}

We present the spectral energy distributions (SEDs) for photometry measured
through different-size apertures for WL16 in Figure~\ref{fig:sed}. The data
have been separated into PSF-fit points (triangles connected by a solid
curve), 8 arcsec aperture photometry points (squares and stars connected by
a dotted curve), and the 45 arcsec resolution HIRES-processed IRAS fluxes
(connected by dot-dashed lines). In the large (8-arcsec) aperture data, the
steep drop in flux approaching the wavelength of 9.7~\microns\ from either
side, was at one time mis-interpreted as a very deep silicate absorption
feature. The depth of this silicate absorption feature is significantly
reduced in the curve connecting just the PSF-fit data points, underlining
the important contribution of PAH feature emission to the larger aperture
photometry.

In the near-infrared, the flux values found within an 8-arcsec aperture
\citep{wil83}, are only slightly higher than the flux values derived
through PSF-fitting, in contrast to the large observed difference between
the PSF--8\arcsec\ aperture photometry values at all mid-infrared
wavelengths, both through filters that include and those that exclude PAH
features (see Figure~\ref{fig:sed}). This behavior is as expected of VSG
emission: providing only a marginal contribution to near-infrared continuum
emission, but being the dominant source of mid-infrared continuum emission.
This result also justifies the use of our near-infrared PSF-fit fluxes to
derive the extinction and intrinsic central source luminosity (see
$\S$\ref{sec:central} below).

\section{Discussion}

\subsection{Nature of the Central Source}\label{sec:central}

The spectral type of the central source of WL16 has recently been
determined from high signal-to-noise, moderate resolution, near-infrared
spectroscopy to lie in the range B8 to A7, corresponding to a photospheric
effective temperature of 8,000~K $\leq T_{eff}\le $ 12,000~K \citep{lr99}.
We combine this constraint on the central object's effective temperature
with other observational data and theoretical pre-main-sequence tracks in a
self-consistent manner to re-determine both the estimated extinction to the
central source and its intrinsic bolometric luminosity, mass, and radius.
The low millimeter flux \citep{and94,mot98}, lack of near-infrared excess
\citep{moo98}, and the fact that UV/optical radiation is able to illuminate
such a large (900 AU) VSG/PAH emitting region in the immediate vicinity of
WL16, all lead to the conclusion that the circumstellar material
surrounding WL16 is optically thin at these wavelengths. Furthermore,
computations of the continuum spectral energy distributions of flared and
geometrically thin accretion disks have shown that high accretion rates
($\dot{M}\geq 10^{-6}$~M$_{\sun }$~yr$^{-1}$) are required to produce
optically thick inner disks around HAeBe stars \citep{har93}. For
comparison, the upper limit for the inner disk accretion rate for WL16 is
$\leq 2\times 10^{-7}$~M$_{\sun }$~yr$^{-1}$\citep{na96a}.

If the circumstellar environment of WL16 is optically thin in the
near-infrared, then the observed extinction along the line-of-sight towards
WL16 must arise in the intervening $\rho$~Oph cloud, which acts as a
foreground screen. We note that this is a radically different physical
source picture than one in which all of the energetic UV/visible source
photons are assumed to be intercepted by the immediately surrounding
circumstellar material, and whose energy is subsequently re-radiated by
this same circumstellar material at longer, infrared wavelengths.

In order to determine the value of the intervening extinction, we fix the
photospheric temperature of the central source at 9000 K. We include only
the near-infrared photometry in the fit, since the near-infrared fluxes are
essentially uncontaminated by any ``classical'' grain or VSG emission
from circumstellar material. We then fit a reddened blackbody (modeling the
extinction with the \citet{mathis} curve for molecular clouds) with only
two free parameters, $A_{V}$ and stellar radius, to determine the
extinction through the cloud.

The resulting extinction, $A_{V}=31$, falls within the range of previously
derived values. A strict upper limit can be derived from the line-of-sight
beam-averaged (55\arcsec\ FWHM) $^{13}$CO observations, from which the
extinction through the entire cloud in the direction of WL16 is estimated
at $A_{V}=70$ mag with a 50\% error (Wilking, priv. comm.). At the other
extreme, allowing for a wide range of intrinsic NIR source colors, a
possible range of $2.0\leq A_{K}\leq 2.9$ was found by \cite{car93}. A
value for $A_{V}=31$ towards WL16 was previously published, derived from
its $H-K$ color by \cite{wil83}. The most recent published determination
for the extinction towards WL16 is an $A_{V}=37$, determined using the NIR
photometry of \cite{gre92}, and assuming an intrinsic A star photosphere
\citep{moo98}.

\begin{figure}[tb]
\includegraphics[width=\columnwidth]{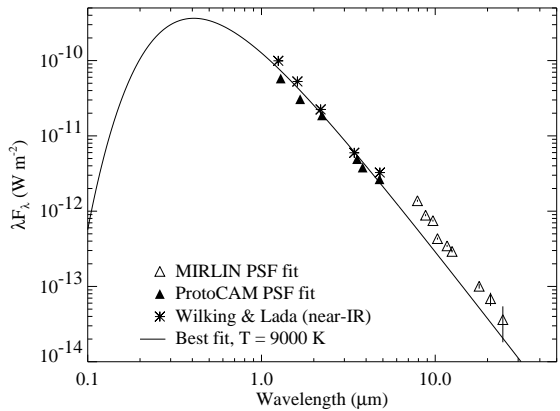}
\caption{The solid curve represents the de-reddened spectral energy
distribution (SED) of a $T_{eff}=9000$~K, $L_{*}=250$~L$_{\sun }$,
$R_{*}=6.5$~R$_{\sun }$ blackbody (photosphere) that represents the central
illuminating source of WL16, seen through an extinction of $A_{V}=31\pm 1$
(see text). Corresponding de-reddened PSF-fit fluxes are indicated by
filled triangles (near-infrared) and open triangles (mid-infrared);
de-reddened fluxes measured through 8-arcsec apertures are indicated by
stars (near-infrared).\label{fig:fit}}
\end{figure}

We argue that our determination of $A_{V}=31\pm 1$ for the extinction
towards WL16 is the most stringent to date. Figure~\ref{fig:fit} shows data
and a model that have all been dereddened for an intervening $A_{V}=31$
screen using the \citet{mathis} extinction curve. The solid curve plotted
in Figure~\ref{fig:fit} shows the resulting spectral energy distribution
for a $T=9000$~K blackbody photosphere with a radius of 6.5 R$_{\sun }$ at
$d=125$~pc. The triangles represent the dereddened PSF-fit data points; the
asterisks represent the corrected near-infrared data from \citet{wil83}.
Note that the residual silicate absorption feature apparent from the solid
curve joining the mid-infrared, PSF-fit photometry in Figure~\ref{fig:sed}
completely disappears from the corresponding extinction-corrected PSF-fit
data plotted in Figure~\ref{fig:fit}. This is especially striking in view
of the fact that the mid-infrared fluxes were not included in the fit used
to determine the extinction to the source. The dereddened, PSF-fit,
mid-infrared data points do lie somewhat above the plotted photospheric
model curve, however. This slight, remaining mid-infrared ``excess'' may
signal the presence of VSGs within $<40$~AU of the central source (within
the PSF). Nevertheless, the goodness of the fit suggests that the residual
``dip'' apparent in the mid-infrared, PSF-fit data of Figure
\ref{fig:sed} is caused by the intervening cloud's silicate absorption, and
not by PAH emission.

Strong constraints on the possible values of the derived extinction can be
made via the two requirements of 1) matching the shape and magnitude of the
dereddened near-infrared flux data to an appropriate blackbody, and 2)
eliminating the residual 9.7~\micron\ silicate absorption feature from the
curve joining the PSF-fit fluxes in Figure \ref{fig:sed}. If $A_{V}$ were
greater than $\sim 32$, the dereddened data would show a silicate emission
feature in the mid-infrared and the near-infrared points would lie on a
curve steeper than a blackbody. For $A_{V}<30$, fits consistent with a
9,000~K blackbody emitter would become problematic. In fact, the derived
$A_{V}$ is relatively insensitive to the assumed temperature: variation
from 6,000 to 12,000~K produces only a 1 magnitude change in $A_{V}$ since
the infrared data are well on the Rayleigh portion of the blackbody curve.
Therefore, we can confidently assert that the range of foreground
extinction falls within $A_{V}=31\pm 1$.

\begin{figure*}[tb]
\includegraphics[width=\textwidth]{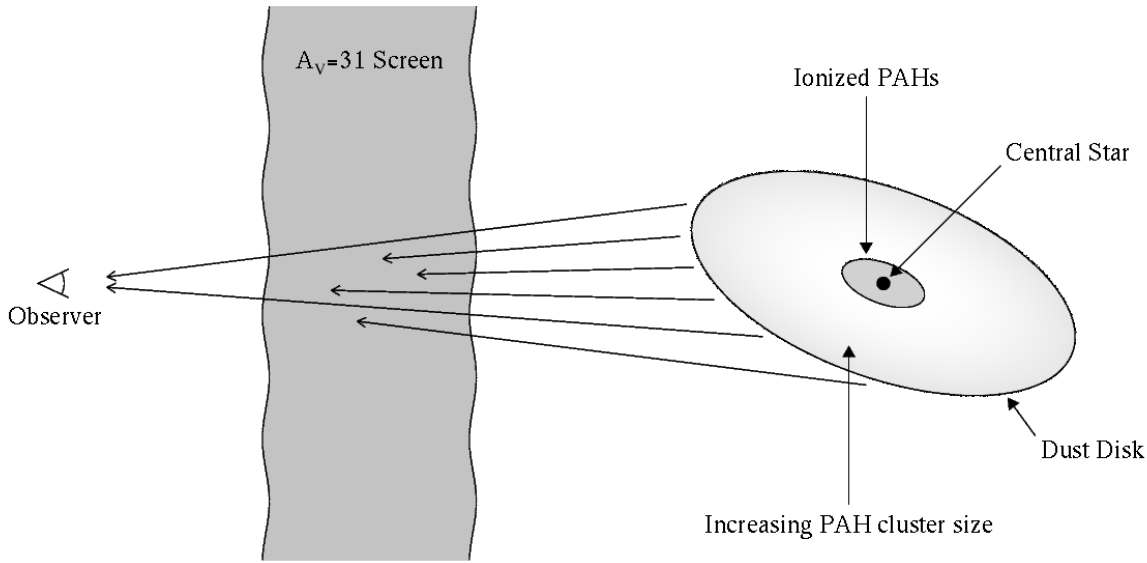}
\caption{A schematic showing the proposed configuration of WL16. The entire
WL16 system (star plus large disk) is viewed through a portion of the $\rho$~Oph cloud, which acts as an $A_{V}=31$ foreground screen. The dust disk
around the star is effectively optically thin, so that UV radiation
propagates through the entire disk. The inner portions of the disk
experience a UV field sufficiently strong to ionize a substantial portion
of the PAHs, while a decrease in the field allows a higher fraction of
large PAHs to exist at the periphery of the disk.\label{fig:scheme}}
\end{figure*}

We can now arrive at a new determination of the bolometric luminosity of
WL16, simply by integrating under the photospheric curve plotted in
Figure~\ref{fig:fit}. We find a bolometric luminosity of 250~L$_{\sun }$
for an assumed distance of 125 pc. This is much higher than previously
published values for the luminosity of WL16, which range from 10 L$_{\sun
}$ to 22 L$_{\sun }$ for $d=160$~pc (\cite{you86,wil89}). This discrepancy
is resolved when one takes into account the very different assumptions that
went into the different luminosity determinations. In the absence of
evidence to the contrary, previous workers had assumed that all the
extinction towards WL16 was local and circumstellar, and that all of the
source photons were absorbed and re-radiated locally at mid- and
far-infrared wavelengths. However, in view of subsequent multi-wavelength
photometry and spectroscopy, the new source picture that emerges is one in
which the extinction is provided by a foreground screen, remote from the
source, so that the absorbed photons are reradiated well outside our (or
anyone else's) beam (see Figure~\ref{fig:scheme} for the schematic cartoon
of the source geometry).

Our new bolometric luminosity determination of 250~L$_{\sun }$, combined
with the spectroscopically determined effective temperature determination
of 9,000~K, allows placement of WL16 on the H-R diagram for the first time.
Comparison of the position of WL16 on this diagram with the
pre-main-sequence tracks of \cite{pal99} yield a source mass of $\sim
4$~M$_{\sun }$ and a source age of $1.9\times 10^{5}$~Yr. The \cite{sie00}
PMS tracks also give a source mass of $\sim 4$~M$_{\sun }$, but an age of
$1.2\times 10^{6}$~Yr. Previous determinations for the central source's
mass in WL16 were made before pre-main-sequence tracks for more massive
objects ($\geq 2.5$~M$_{\sun }$) were available, coupled with an erroneous
underestimate (10--20 L$_{\sun }$) of the central object's true luminosity.

\subsection{A Mid-IR Emitting Disk in WL16}

Previous mid-infrared imaging of WL16 shows an extended emission region
elongated along an approximately NE--SW axis \citep{deu95,dev98,moo98}.
This emission has variously been interpreted as originating from a nebula,
a torus, a bipolar cavity, or a disk. We consider a ``nebula'' (meaning
an \ion{H}{2} region) to be unlikely given the lack of narrow spectral
features in the near- or mid-infrared, and the fact that the central star
has a much later spectral type than B3 \citep{lr99}. It is also not a
reflection nebula given that scattering is inefficient in the mid-infrared,
and the density necessary to show any scattering would require an optically
thick disk; we have already argued that WL16's emission is optically thin.
A torus is unlikely since the intensity profile does not indicate a
``hole'' in the inner regions of the disk, and a torus would be difficult
to maintain physically. A bipolar cavity can be ruled out since the outflow
to create such a cavity was searched for, but not found \citep{cab91}. We
propose that the most plausible interpretation of our multi-wavelength
imaging dataset favors the notion that the mid-infrared emission originates
from a geometrically thin, equatorial disk.

Under this assumption for the source geometry, the minor-to-major axis
ratio is found to be $0.466\pm 0.007$, measured from the images as observed
in the four filters that include the PAH emission features. This ratio
corresponds to an inclination angle of $62.2^{\circ }\pm 0.4^{\circ }$ of
the disk to our line-of-sight, if we assume the disk to be intrinsically
circular and to appear elongated only because of projection effects. This
value agrees well with the inclination angle ($i=60^{\circ }$) deduced for
the inner disk ($R\leq 30$~R$_{\sun }$) of WL16 from kinematic modeling of
the hot gas ($T=5000$~K) emitting in the near-infrared CO lines
\citep{den91,car93,cha93,cha95}.

\subsection{Properties of the Mid-IR Emitting Disk}

It has already been suggested that the mid-IR emitting dust surrounding
WL16 consists of at least two components, aromatic hydrocarbons and VSGs
\citep{moo98}. The present observations strongly support this suggestion.
The disk of WL16 is easily visible through all the mid-infrared filters,
not just those which contain PAH emission features, although the disk does
emit most strongly through those filters which encompass PAHs
(Figure~\ref{fig:images}). Similar results were found by \citet{moo98},
upon comparison of their images of WL16 through filters which contain PAH
emission features at 8.2 and 11.3~\microns\ with their 10.2~\micron\
continuum image, which is PAH-feature-free. Our results extend this finding
to a greater number of wavelengths.

Simple modeling of PAH and VSG emission in the mid-infrared from spherical
distributions of optically thin dust surrounding an 8,000~K (12,000~K)
central source, suggests that 1\% (3.5\%) of the starlight would be
absorbed by PAHs and 15\% (20\%) by VSGs. However, these values for the
PAHs are highly dependent on the assumed absorption cross-sections, and
could be as high as 16.6\% (17\%) for an 8,000~K (12,000~K) central object
\citep{nat95}. For WL16, we find that $\sim $ 1\% of the 9,000~K central
source's luminosity, or 2.6~L$_{\sun }$, are emitted by the PAHs and VSGs.
At the very least, this low value, relative to the models, rules out
spherical distributions of PAHs and VSGs around the source. For comparison,
previous authors have estimated $\sim 0.6$~L$_{\sun }$ to be contributed by
the 7.7--13 \micron\ PAH band emission within a 5\arcsec\ beam
\citep{han92}, scaled to $d=125$~pc, and 1.2~L$_{\sun }$, integrated over
the entire disk extent \citep{moo98}.

\subsubsection{Very Small Grain (VSG) Constituent}

Figure~\ref{fig:images} shows the extended appearance of WL16 at 9.7, 10.3,
17.9, 20.8, and 24.5~\microns; passbands which exclude PAH feature
emission. Since neither PAHs nor the ``standard'' MRN grains can account
for this emission, as we argue below, the continuum emitter responsible for
this extended mid-infrared emission in the disk of WL16 must be in the form
of the so-called very small grains (VSG's) \citep{nat95,dev98,moo98}. It
has been shown that such VSGs surrounding HAeBe stars are excited by a much
broader spectrum of photons than the PAHs and can reprocess stellar
radiation into the mid-IR \citep{nat93}.

\begin{figure}[!ht]
\includegraphics[width=\columnwidth]{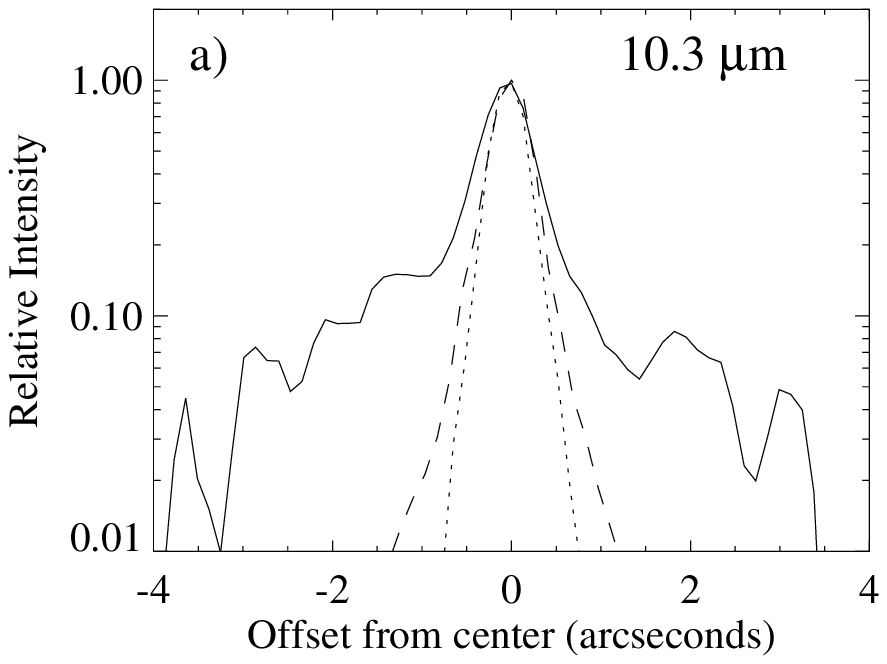}
\includegraphics[width=\columnwidth]{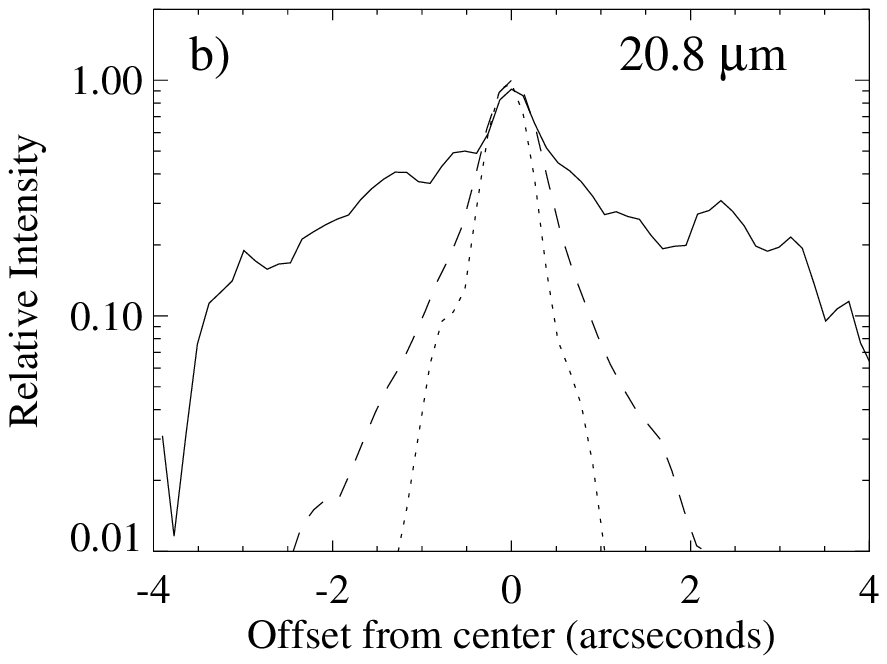}
\caption{Plots of the observed continuum intensity profile of the disk of
WL16 along its major axis compared with models of a disk composed only of
classical interstellar grains. Panel a) shows the data and model for 10.3
\microns; panel b) for 20.8 \microns. The solid curves represent the
observed profiles; the dashed curves show the interstellar-grain-only disk
model, and the dotted curves are the observed PSFs. While the disk models
are clearly more extended than the observed PSFs, they underpredict the
observed flux from WL16's disk by an order of magnitude or more at radii
greater than an arcsecond, and their steep slope is incompatible with the
broad observed profiles. The presence of VSGs is required to account for
the large spatial extent of non-PAH emission.
\label{fig:mrn}}
\end{figure}

We have already pointed out (see $\S$\ref{sec:central}) that the systematic
mid-infrared excess flux that remains after de-reddening the PSF-fit
photometry requires the presence of VSG's on scales $\leq 40$~AU (see
Figure \ref{fig:fit}). Similarly, emission from VSGs is required to explain
the large spatial extent of WL16's disk at continuum wavelengths ($R \sim
450$~AU). This is quantitatively illustrated by Figure \ref{fig:mrn}, where
intensity cross-cuts of a model disk, consisting of regular ISM grains, are
compared with the actual, observed intensity profiles of WL16's disk. The
model disk intensity profiles were kindly provided by Dr.~Barbara Whitney,
using a 3-D Monte Carlo code described in \citet{wood}, and includes both
scattering and emission by standard interstellar grains \citep{kim} at
mid-infrared wavelengths. The model shown is for a 0.01 M$_\sun$ flared
disk (the maximum disk mass allowed by millimeter observations), inclined
at 60$^{\circ}$ to our line-of-sight, at a distance of 125 pc, illuminated
by a central source of 250 L$_\sun$ and $T_{eff}=9000$~K. The disk density
structure is as described in \citet{wood}, except the inner radius is set
by the dust destruction radius for a $T_{eff}= 9000$~K central source, and
the disk's outer radius is determined by the observations presented here.
Note that emission from standard, interstellar grains at large radii from
the central source would be undetectable, according to the models, in stark
contrast to the observed emission. Standard, interstellar grains cannot,
therefore, be responsible for the appearance of WL16's disk at large radii
through the filters which exclude PAH emission.

Additional qualitative proof of this assertion is provided in
Figure~\ref{fig:cuts}. A comparison of the 9.7, 10.3, 17.9, 20.8, and 24.5
\micron\ flux profiles (none of which should have significant PAH emission)
with the fiducial 11.7~\micron\ profile suggests a rather constant ratio as
a function of distance from the center, despite any differences in the
absolute surface brightness values. This also holds true in the fully
two-dimensional case, as well, where ratios of any pair of images,
including the 10.3 \micron/20.8 \micron\ ratio image (not shown), are
generally very flat outside of the ionized PAH core. This behavior is
consistent with the emission expected from VSGs, but is inconsistent with
the emission properties expected from ``classical'' grains. It also
suggests that the PAHs and VSGs are well mixed.

\subsubsection{PAH Ionization, Hydrogenation, and Size Effects in the WL16 Disk}

\label{sec:pah}

The color ratio images presented in Figure~\ref{fig:gradients} can give
some qualitative indication as to the charge state, amount of
hydrogenation, and size of the emitting PAH molecules. Quantum mechanical
modeling and laboratory studies of PAH emission have shown that the charge
state of the PAH has an important effect on the emitted IR spectrum. In
particular, the intensities of the C$-$C stretching and C$-$H in-plane
bending modes, which fall in the 6--9~\micron\ range, are generally an
order of magnitude stronger in PAH cations (singly ionized PAHs) than in
neutral PAHs \citep{job96,lan96,hu95a,hu95b}. Viewed in this context, it is
therefore not surprising that the 7.9/8.8~\micron\ flux ratio shown in the
top panels of Figure~\ref{fig:gradients} is essentially constant over the
entire emitting region of the WL16 disk, since both of these features (at
7.7 and 8.6~\microns) originate from positively charged PAHs.

\citet{dev98} found a central peaking of the continuum-subtracted
8.6/11.3~\micron\ feature ratios in the disk of WL16. We confirm this in
the central, left-hand panel of Figure~\ref{fig:gradients}, which shows the
8.8/11.7~\micron\ flux ratio image. In contrast to the featurelessness of
the 7.9/8.8~\micron\ ratio image, a distinct enhancement within the central
$\pm $1.5\arcsec\ core is observed (see also Figure~\ref{fig:cuts}).

Recent quantum mechanical calculations show that the 8.6~\micron\ feature
originates mostly from C$-$H in-plane bending modes of positively charged
PAHs, whereas the 11.3~\micron\ feature originates from C$-$H out-of-plane
bending modes of predominantly neutral and anionic PAHs \citep{bak01}. The
observed contrast in the 8.8/11.7~\micron\ flux ratio image and cross-cut
of Figure~\ref{fig:gradients} may be interpreted as a change in charge
state of the PAHs from positively ionized in the central regions to neutral
in the outer regions \citep{job96,dev98}.

The physical conditions required for the 11.3 $\mu$m feature to originate
mainly from neutral, and not positively charged, PAHs, are expressed by the
constraint, $G_{0}/n_{e}\leq 10^{3}$, where $G_{0}$ is the UV radiation
field (6--13.6~eV) in units of $1.6\times 10^{-3}$~ergs~cm$^{-2}$~s$^{-1}$
and $n_{e}$ is the electron density in cm$^{-3}$ \citep{bak01}. Of course,
the central Herbig Ae star is the photo-ionizing and UV source for the WL16
disk. The calculated UV radiation field in the WL16 disk at different
distances from the central source is: $G_{0}=1.2\times 10^{5}$ at
0\arcsec--1\arcsec, $2\times 10^{4}$ at 1\arcsec--2\arcsec, $4.8\times
10^{3}$ at 2\arcsec--4\arcsec, and $1.6\times 10^{3}$ for distances at
4\arcsec--6\arcsec\ \citep{dev98}. Combining these values of the UV flux
with the 11.3 $\mu$m feature neutrality constraint, leads to derived
electron densities in the WL16 disk decreasing from $n_{e}\sim
100$~cm$^{-3}$ in the central regions to $n_{e}\sim 10$~cm$^{-3}$ in the
outer regions.

The bottom panels of Figure~\ref{fig:gradients} show the 12.5/\-11.7~\micron\
flux ratios. The most interesting feature in this image is the gradual
enhancement of the 12.5/11.7~\micron\ ratio towards the disk edges. Two
possible explanations for this enhancement are increasing molecular size
and increasing hydrogenation towards the disk edges.

It has been shown that partially hydrogenated small ($<25$ C~atoms) PAHs,
as well as large PAHs with 1, 2, or 3 adjacent H~atoms can contribute to
the 11.3~\micron\ feature \citep{coh85,mor83}. However, only relatively
large PAHs, with 1, 2, or 3 adjacent H~atoms, are needed to produce the
12.7~\micron\ feature \citep{dul81}. Quantum mechanical calculations show
that the 12.7~\micron\ PAH feature strength increases with molecule size
\citep{bak01}. Therefore, one possible interpretation for the enhancement
of the 12.5/11.7~\micron\ flux ratio at the disk's edges is the growth of
larger PAHs ($\geq 50$--80 C atoms) at greater distances from the central
ionizing source.

Alternatively, since the 11.3~\micron\ feature originates from PAHs with
isolated H~atoms \citep{coh85}, whereas the 12.7~\micron\ feature requires
two adjacent hydrogen atoms on the aromatic ring \citep{ver96}, the slight
increase in the 12.5/11.7~\micron\ ratio apparent towards the edges of the
WL16 disk might be due to increasing hydrogenation. Increasing
hydrogenation may take place due to lower collision rates with electrons
that would knock away H~atoms due to lower $n_{e}$ at the disk edges and/or
lower UV flux to dissociate H~atoms from PAHs.

\subsubsection{Disk Asymmetry}

The intensity profiles along the disk's major axis presented in
Figure~\ref{fig:cuts} show the presence of an asymmetry in emission
strength with respect to mirror reflection about the central position at
all observed wavelengths. This asymmetry occurs at angular separations
between 1\arcsec--2\farcs5 from the central position, corresponding to
radii of 125~AU $\leq r\leq $ 300~AU. Since this asymmetry is observed at
all wavelengths, it is likely due to either a column density enhancement in
one side of the disk or a gap in the other. We do not have enough
information to distinguish between these two possibilities.

Although it is tempting to invoke the presence of embedded planets to
account for the observed disk asymmetry due to gravitational perturbations,
the large radii (125~AU $\leq r\leq $ 300~AU) at which this asymmetry is
observed conflicts with all current theories which predict $\sim 20$~AU as
a typical outer boundary for planet formation from an accretion disk.
Furthermore, any planetary time formation timescale at such large radii
would exceed the inferred age of the central star. Nevertheless, objects a
few meters to hundreds of meters in size could have had time to form at
these radii \citep{ken99}.

\subsubsection{Disk Lifetime}

One intriguing problem is the question of how the PAH/VSG disk we observe
is replenished, since even at the disk's extremities, the timescales for
blowing small particles out of the system by radiation pressure are only
years. Condensation from the gas phase is highly unlikely, given the low
gas densities throughout most of the disk, and the high activation energies
required for PAH synthesis, which leads most researchers in the field to
conjecture that PAHs are formed on the surface of grains. This leads us to
the suggestion that the most likely scenario for replenishment of the
PAHs/VSGs in the disk of WL16 is from the breakup of larger bodies, such as
comets or Kuiper belt objects, or from UV or shock processing of larger
grains as has recently been suggested by \citet{dul00}. The presence of
PAHs in a possible planet-forming system has interesting implications for
the potential seeding of planetary systems with organic materials, and is
also consonant with the discovery of PAHs in meteorites.

\section{Conclusions}

We have presented diffraction-limited images and photometry of the disk
surrounding the Herbig Ae star, WL16, at nine mid-infrared wavelengths, as
well as photometry of the central star at five additional near-infrared
wavelengths. From our newly derived spectral energy distribution for this
source, we find that the central star and its disk are observed through a
foreground screen which provides an extinction of $A_{V}=31\pm 1$
magnitudes. We find that the star itself has a bolometric luminosity of
250~L$_{\sun }$ (at an assumed 125 pc distance), and the disk contributes
$\approx $ 1\% due to very small grain (VSG) and polycyclic aromatic
hydrocarbon (PAH) feature emission.

The unprecedented spatial resolution of the mid-infrared images presented
here allows us to confirm the presence of ionized PAHs in the central disk
regions, and we report the discovery of a population of larger ($\geq
50$--80 C atoms) PAHs and/or more hydrogenated PAHs at the disk's
periphery. We confirm that a population of graphitic VSGs is required to
account for the observed large disk extent through the five mid-infrared
filters which exclude the wavelengths of the PAH emission features. These
VSGs are also required to explain the observed mid-infrared excess emission
within the unresolved inner core ($\leq 0.3$\arcsec) of the WL16 images.

The disk size is found to be 7\arcsec$\times $3.5\arcsec, corresponding to
a de-projected disk diameter of 900~AU at the source. The disk inclination
is found to be $62^{\circ }\pm 2^{\circ }$ to our line of sight at P.A.
60$^{\circ }$. We find an asymmetry in the disk, at all observed mid-IR
wavelengths, at 1\arcsec--2\farcs5 from the central source. Finally,
continuous replenishment of disk material, possibly from collisions of
larger parent bodies, is required to maintain the PAH/VSG disk, which would
otherwise be destroyed by blowout by radiation pressure on timescales of
years.

\acknowledgements

We wish to thank Dr.~Emma Bakes, Dr.~Timothy Brooke, Prof.~Lynne K.
Deutsch, Dr.~Tom Greene, Dr.~Martha Hanner, Dr.~Antonella Natta,
Dr.~Deborah Padgett, Dr.~Sue Terebey, Dr.~Michael Werner, and Prof.~Bruce
Wilking for helpful discussions during the preparation of this paper.
Dr.~Barbara Whitney kindly provided realistic disk models consisting of
interstellar grains for direct comparison with the observations presented
in this paper. MR thanks Dr.~Fred Chaffee and the entire Keck Observatory
staff for their enthusiasm, patience, and assistance in making it possible
to use MIRLIN on the Keck II telescope.

Portions of this work were carried out at the Jet Propulsion Laboratory,
California Institute of Technology, under contract with the National
Aeronautics and Space Administration. Development of MIRLIN was supported
by the JPL Director's Discretionary Fund and its continued operation is
funded by an SR+T award from NASA's Office of Space Science. MB gratefully
acknowledges financial suppport from grants NSF AST97-53229, NSF 0096087
(CAREER), NSF AST02-06146, and from NASA's Long-Term Space Astrophysics
Research Program, NAG5 8412, which made her contributions to this work
possible.

\end{document}